\documentclass[aps,prx,superscriptaddress,twocolumn]{revtex4-1}

\usepackage[utf8]{inputenc}
\usepackage{amsmath} 
\usepackage{amssymb} 
\usepackage{amsfonts} 
\usepackage{graphicx}   
\usepackage{verbatim}  
\usepackage{color}     
\usepackage{hyperref}   
\usepackage{gensymb}
\usepackage{color}
\usepackage{multirow}
\usepackage{xfrac}
\begin{document}

\title{Phase Separation in the Vicinity of Fermi Surface Hot Spots}
%Intrinsic Phase Separation in Electron-Doped 1$T$-TiSe$_2$ Close to the Excitonic Instability
%Phase Separation in Electron-Doped 1$T$-TiSe$_2$ with Imperfect Crossing at the Fermi Surface
%hase Separation Driven by Imperfect Crossing Points at the Fermi Surface of Electron-Doped 1$T$-TiSe$_2$
%with Fermi Surface Lukewarm Spots$
%Origin of Electronic Phase Separation in Electron-Doped 1$T$-TiSe$_2$: Imperfect Crossing of the Electron and Hole Fermi Surface Sheets
\author{T. Jaouen}
\altaffiliation{Corresponding author.\\ thomas.jaouen@unifr.ch}
\affiliation{D{\'e}partement de Physique and Fribourg Center for Nanomaterials, Universit{\'e} de Fribourg, CH-1700 Fribourg, Switzerland}

\author{B. Hildebrand}
\affiliation{D{\'e}partement de Physique and Fribourg Center for Nanomaterials, Universit{\'e} de Fribourg, CH-1700 Fribourg, Switzerland}

\author{M.-L. Mottas}
\affiliation{D{\'e}partement de Physique and Fribourg Center for Nanomaterials, Universit{\'e} de Fribourg, CH-1700 Fribourg, Switzerland}

\author{M. Di Giovannantonio}
\affiliation{Empa, Swiss Federal Laboratories for Materials Science and Technology, nanotech@surfaces Laboratory, 8600 Dübendorf, Switzerland}

\author{P. Ruffieux}
\affiliation{Empa, Swiss Federal Laboratories for Materials Science and Technology, nanotech@surfaces Laboratory, 8600 Dübendorf, Switzerland}

\author{M. Rumo}
\affiliation{D{\'e}partement de Physique and Fribourg Center for Nanomaterials, Universit{\'e} de Fribourg, CH-1700 Fribourg, Switzerland}

\author{C. W. Nicholson}
\affiliation{D{\'e}partement de Physique and Fribourg Center for Nanomaterials, Universit{\'e} de Fribourg, CH-1700 Fribourg, Switzerland}

\author{E. Razzoli}
\affiliation{Quantum Matter Institute, University of British Columbia, Vancouver, BC, Canada V6T 1Z4}
\affiliation{Department of Physics and Astronomy, University of British Columbia, Vancouver, BC, Canada V6T 1Z1}

\author{C. Barreteau}
\affiliation{Department of Quantum Matter Physics, University of Geneva, 24 Quai Ernest-Ansermet, 1211 Geneva 4, Switzerland}

\author{A. Ubaldini}
\affiliation{Department of Quantum Matter Physics, University of Geneva, 24 Quai Ernest-Ansermet, 1211 Geneva 4, Switzerland}

\author{E. Giannini}
\affiliation{Department of Quantum Matter Physics, University of Geneva, 24 Quai Ernest-Ansermet, 1211 Geneva 4, Switzerland}

\author{F. Vanini}
\affiliation{D{\'e}partement de Physique and Fribourg Center for Nanomaterials, Universit{\'e} de Fribourg, CH-1700 Fribourg, Switzerland}

\author{H. Beck}
\affiliation{D{\'e}partement de Physique and Fribourg Center for Nanomaterials, Universit{\'e} de Fribourg, CH-1700 Fribourg, Switzerland}

\author{C. Monney}
\affiliation{D{\'e}partement de Physique and Fribourg Center for Nanomaterials, Universit{\'e} de Fribourg, CH-1700 Fribourg, Switzerland}

\author{P. Aebi}
\affiliation{D{\'e}partement de Physique and Fribourg Center for Nanomaterials, Universit{\'e} de Fribourg, CH-1700 Fribourg, Switzerland}

\begin{abstract}

Spatially inhomogeneous electronic states are expected to be key ingredients for the emergence of superconducting phases in quantum materials hosting charge-density-waves (CDWs). Prototypical materials are transition-metal dichalcogenides (TMDCs) and among them, 1$T$-TiSe$_2$ exhibiting intertwined CDW and superconducting states under Cu intercalation, pressure or electrical gating. Although it has been recently proposed that the emergence of superconductivity relates to CDW fluctuations and the development of spatial inhomogeneities in the CDW order, the fundamental mechanism underlying such a phase separation (PS) is still missing. Using angle-resolved photoemission spectroscopy and variable-temperature scanning tunneling microscopy, we report on the phase diagram of the CDW in 1$T$-TiSe$_2$ as a function of Ti self-doping, an overlooked degree of freedom inducing CDW texturing. We find an intrinsic tendency towards electronic PS in the vicinity of Fermi surface (FS) "hot spots", i.e. locations with band crossings close to, but not at the Fermi level. We therefore demonstrate an intimate relationship between the FS topology and the emergence of spatially textured electronic phases which is expected to be generalizable to many doped CDW compounds. 

\end{abstract}

\date{\today}
\maketitle

\begin{figure*}[t]
\includegraphics[width=1\textwidth]{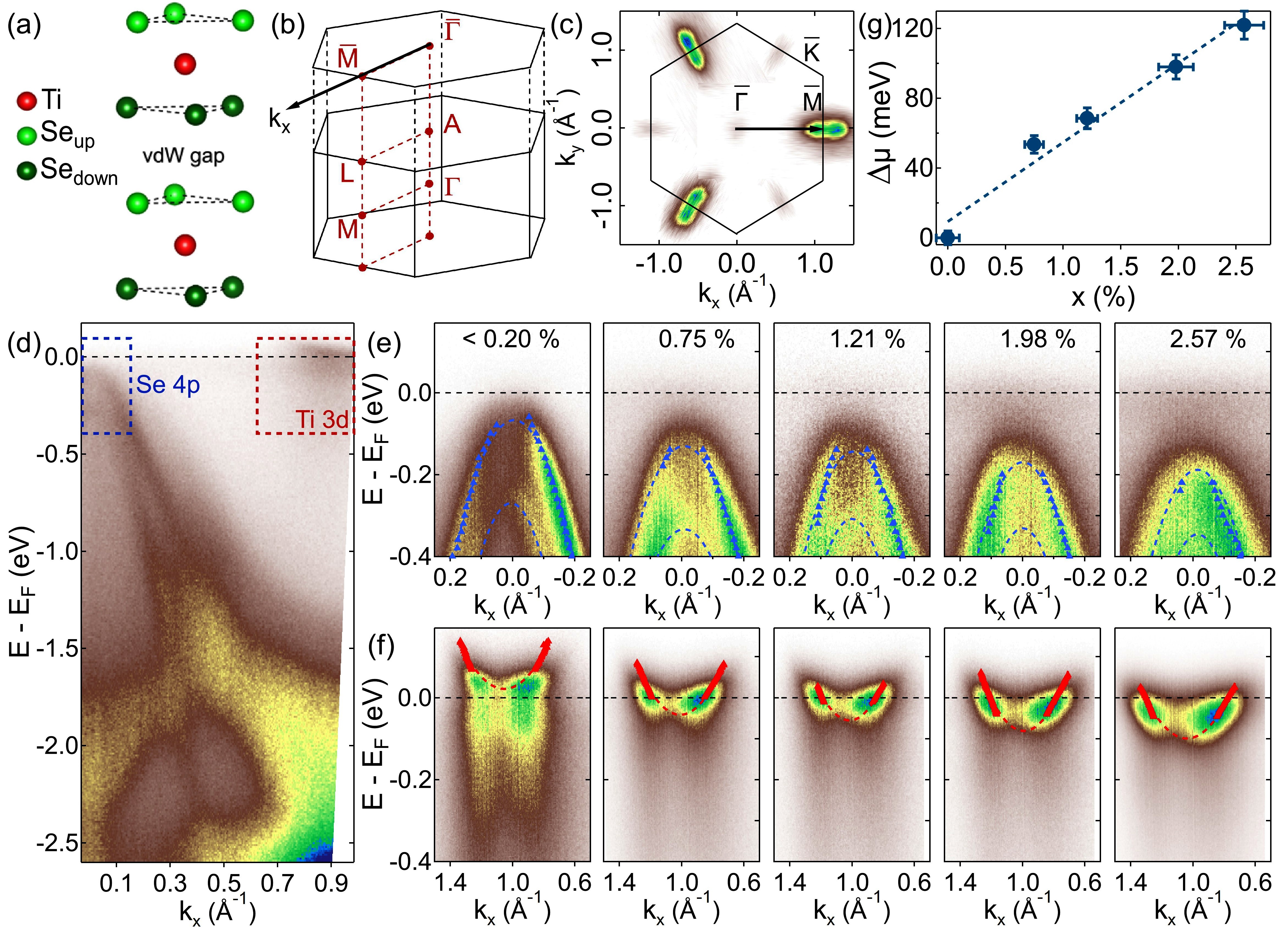}
\caption{(a) Side view of two 1$T$-TiSe$_2$ layers separated by a van der Waals gap. Se$_{up}$ and Se$_{down}$ respectively refer to atoms of the top and bottom Se atomic planes of a 1$T$-TiSe$_2$ layer. (b) Three-dimensionnal and surface Brillouin zone (BZ) of 1$T$-TiSe$_2$. (c) RT Fermi surface (FS) of pristine 1$T$-TiSe$_2$ as measured using He-I radiation ($h\nu$=21.22 eV). The $k_x$ axis depicted on (b) and (c) by the black arrow is the $\bar{\mathrm{\Gamma}}$-$\bar{M}$ direction of the surface BZ. (d) Large-energy scale ARPES spectrum of pristine along the $k_y$=0 \AA$^{-1}$ cut of the FS. The blue and red-dashed boxes respectively indicate the energy and $k_x$ windows shown in (e) and (f) for the hole and electron pockets. (e)-(f), RT ARPES intensity maps, as false color plots (blue colors represent strong intensity) as a function of $x$ at the $\bar{\mathrm{\Gamma}}$ and $\bar{M}$ points of the surface BZ. He-I radiation is used and the doping concentrations, $x$, are $<$ 0.20 $\%$ , 0.75$\pm$0.10 $\%$, 1.21$\pm$0.14 $\%$, 1.98$\pm$0.15 $\%$, and 2.57$\pm$0.22 $\%$ from left to right. Blue, respectively red, dashed lines and triangles on (e) and (f) correspond to parabolic fits of the maxima of the MDC curves (triangles). (g) Linear dependence of the experimental chemical-potential shift $\Delta \mu$ with $x$ as extracted from the evolution of the hole and electron band extrema.}\label{fig1}
\end{figure*}
 
\begin{figure*}[t]
\includegraphics[width=1\textwidth]{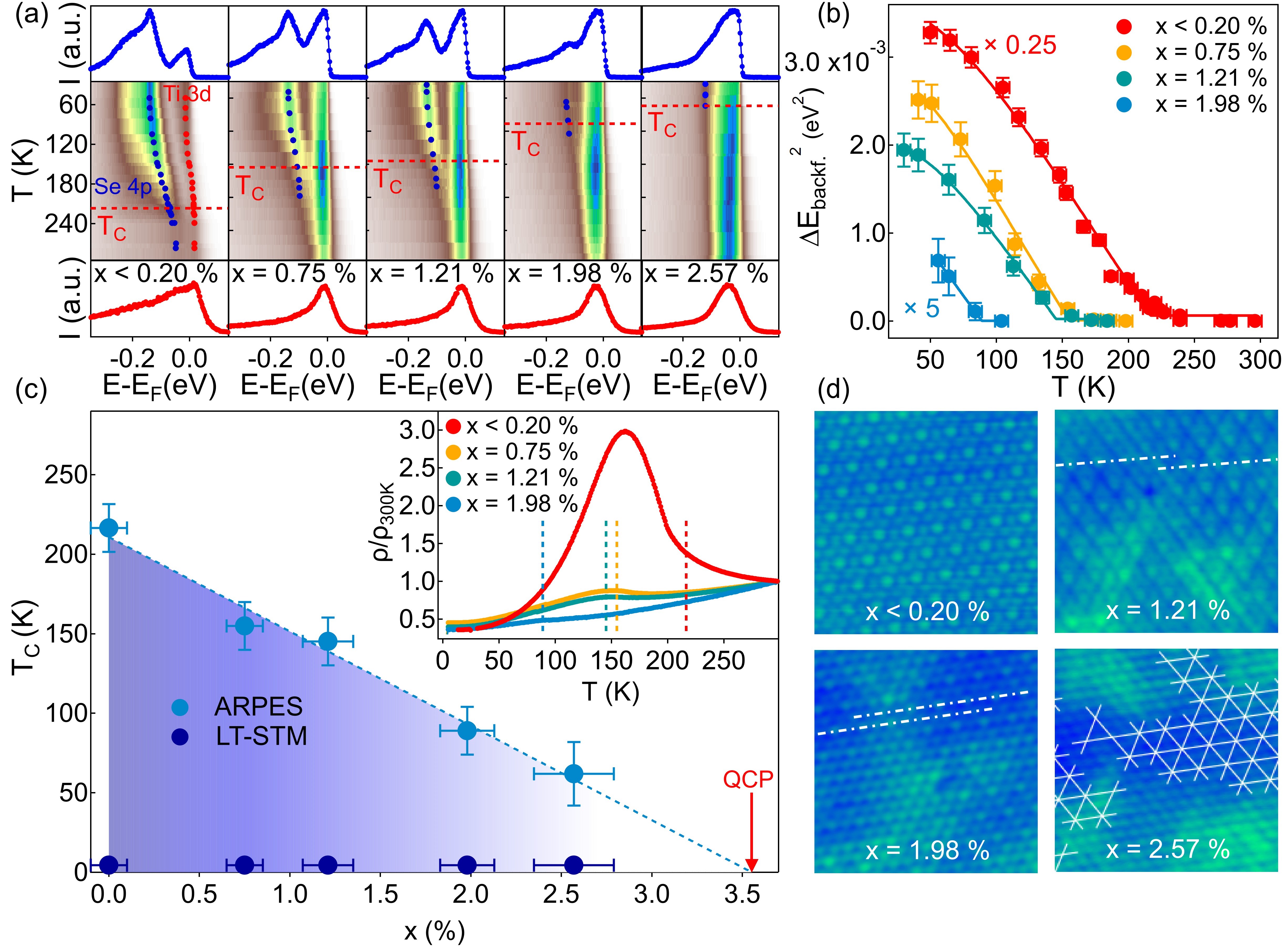}
\caption{ (a) $T$-dependent EDC maps measured at $\bar{M}$ as a function of Ti self-doping (from left to right). For each doping, lower and upper panels respectively show the RT and LT EDCs. The red-dashed lines locate the critical CDW transition temperatures ($T_c$) extracted from the energy shift of the Se 4$p$ backfolded band (blue dots) ${\Delta}E_{backf}$ as a function of $T$ using a BCS-type gap equation as shown in (b). (c) $T$-$x$ phase diagram of Ti-doped 1$T$-TiSe$_2$. The CDW transition temperature $T_c$ is linearly decreasing with $x$ towards a QCP. The inset shows $T$-dependent resistivity curves measured on the pristine (red curve) and Ti-doped crystals. The curves are normalized to the resistivity measured at RT and the vertical dashed-lines show the location of $T_c$ for each doping as determined using ARPES. (d) 5.5 $\times$ 5.5 nm$^2$ STM images taken at 4.5 K (see dark blue points on (c)) of pristine ($x$ $<$0.20 $\%$) and Ti-doped crystals ($x$=1.21 $\%$, 1.98 $\%$, and 2.57 $\%$). $I$=0.2 nA, $V_{\text{bias}}$=150 mV (pristine), 50 mV (1.21 $\%$), 100 mV (1.98 $\%$), and -50 mV (2.57 $\%$). Added meshes on the $x$=2.57 $\%$ STM image and white-dashed lines on the $x$=1.21 $\%$ and 1.98 $\%$ ones highlight CDW domains and phase shifts between them, respectively.}\label{fig2}
\end{figure*}

\paragraph*{Introduction}

It is well established that electronic phase separation (PS) at the nanometer scale can emerge from competing short-range and long-range interactions and is accompanied by charge inhomogeneity especially in doped systems \cite{Li2015, Mamin2018}. This has been studied in the context of so-called Coulomb-frustration where the competition between the long-range Coulomb interaction, enhanced due to excess charge, and surface energy effects control the characteristic size of the phase-separated pattern \cite{Ortix2008}. PS has been demonstrated to lie, for example, at the heart of the magnetoresistance manganite \cite{Dagotto2005}, and high-$T_c$ cuprate physics \cite{Campi2015}. Another important materials class exhibiting phase diagrams including intertwined charge-density-waves (CDWs), Mott states and superconductivity is the family of layered transition-metal dichalcogenides (TMDCs) \cite{Morosan2006a, Sipos2008, Kusmartseva2009a}. 

Among them, 1$T$-TiSe$_2$ is a prototypical material with a 2 $\times$ 2 $\times$ 2 commensurate CDW (CCDW) occuring at $\sim$200 K and hosting superconductivity under pressure, Cu-doping or electrical gating \cite{Kusmartseva2009a, Morosan2006a, Li2015}. The emergence of superconductivity is currently thought to relate to a change in the nature of the CDW from CCDW to incommensurate (ICDW) \cite {Kogar2017, Joe2014}, passing through a nearly-commensurate intermediate regime (NCCDW) constituted of CCDW domains separated by atomically sharp domain walls (DWs), or discommensurations, hosting excess electrons \cite{Chen2019, Yan2017, Novello2017}.

Intercalation of Ti dopants in the van der Waals (vdW) gap of the 1$T$-TiSe$_2$ structure [see Fig. \ref{fig1}(a)] is known to occur depending on the crystal growth temperature \cite{salvo1976}, and is found even in the best available crystals. Although not superconducting, Ti self-doped 1$T$-TiSe$_2$ crystals have also shown to exhibit a spatially inhomogeneous landscape ultimately leading to nanoscale PS between CDW domains and non-distorted 1 $\times$ 1 phase \cite{Hildebrand2016, Hildebrand2017, Hildebrand2018}. As we will show, most of the debate surrounding the semimetallic or semiconducting character of the 1$T$-TiSe$_2$ in the high-temperature phase and the underlying origin of the CDW phase transition \cite{Rossnagel2011, Kogar2017b} are related to the electron-doping nature of such defects. This, in addition to the fluctuating nature of the CDW at room-temperature (RT) made the Fermi surface (FS) topology in the normal state of \textit{pristine} 1$T$-TiSe$_2$ elusive until now.

Combining angle-resolved photoemission spectroscopy (ARPES), low- and variable-temperature scanning tunneling microscopy (LT-, VT-STM), we study 1$T$-TiSe$_2$ crystals containing low concentrations of native Ti intercalants and focus on the temperature ($T$)-driven and concentration ($x$)-induced CDW transition and suppression. We demonstrate that in pristine 1$T$-TiSe$_2$ the CDW phase transition occurs at a temperature where the chemical potential matches the electron and hole band crossings (defined as FS "hot spots") in the \textit{hidden} semimetallic normal state. In Ti-intercalated samples, the electron-hole band crossings move below the Fermi level due to rigid electron doping. For locally recovering FS hot spots, the system undergoes PS accompanied by charge inhomogeneities. Finally, VT-STM measurements reveal the real-space emergence of the CDW upon temperature lowering and uncover its inhomogeneous nature. The demonstrated PS in a wide region of the ($x$-$T$) phase diagram is shown to primarily rely on the $T$- and $x$-dependences of the chemical potential, i.e. of the FS topology. Our study reinforces the key role of the FS topology for driving not only the CDW instability in 1$T$-TiSe$_2$ but also phase-separated states that emerge in a large class of TMDCs upon doping or pressure \cite{Qiao2017, Liu2016, Ang2012, Xu2010, Sipos2008, Wu2007}.

\paragraph*{Rigid doping by Ti intercalation}

In the 1 $\times$ 1 $\times$ 1 normal state, the 1$T$-TiSe$_2$ low-energy electronic states consist of a Se 4$p$ hole pocket at $\mathrm{\Gamma}$ and Ti 3$d$ electron pockets at the three equivalent $L$ points of the three-dimensional Brillouin zone (BZ) [see Fig. \ref{fig1}(b)]. At the CDW transition, the $\mathrm{\Gamma}$ and the $L$ points are connected by the three new reciprocal lattice $q$-vectors corresponding to the doubling of the lattice periodicity. Figure \ref{fig1}(c) and \ref{fig1}(d) respectively show the RT FS and the corresponding $k_y$=0 \AA$^{-1}$ cut for pristine 1$T$-TiSe$_2$. In a free-electron final-state picture He-I ARPES ($h \nu$=21.22 eV) mainly probes the 1$T$-TiSe$_2$ initial states close to the $A$-$L$ high-symmetry line of the BZ [see Fig. \ref{fig1}(b)] \cite{Pillo2000}. Details of the experimental method are given in Appendix A and procedures for the ARPES data analysis are described elsewhere \cite{Mottas2019}. For pristine 1$T$-TiSe$_2$, the hole band at $A$ [Fig. \ref{fig1}(e), left panel] is below the Fermi level ($E_F$) (with the top of the band at 67$\pm$2 meV binding energy) and the Ti 3$d$ electron pocket at $L$ [Fig. \ref{fig1}(f), left panel] is only thermally occupied and lies 21 meV above $E_F$. The observed spectral weight centered at $\sim$80 meV below the Ti 3$d$ band manifests CDW correlations already at RT \cite{Monney2012b}. In fact, the $q$-vectors of the 2 $\times$ 2 $\times$ 2 CDW connect the $\Gamma$ and $L$ points of the BZ. The CDW therefore manifests itself via umklapp, or backfolding, of the Se 4$p$ hole band at $\Gamma$ to $L$ [visible in Fig. \ref{fig1}(f), left].  

Figures \ref{fig1}(e), (f) show RT-ARPES intensity map as a function of Ti self-doping (from left to right) at the $A$ and $L$ points of the BZ, respectively. We notice that our \textit{first} Ti doping of 0.75$\pm$0.1 $\%$ already moves the electron pocket below $E_F$ thereby occupying it. The evolution of the hole and electron band extrema extracted from the momentum distribution curves (MDC) fitted as a function of doping (overlaid on the data) further shows that a rigid band model of electron doping applies with the chemical potential $\mu$, which is linearly raised up by 122 meV between pristine and $x$=2.57 $\%$ [see Fig. \ref{fig1}(g)]. 

\paragraph*{Phase diagram}

Focusing on the impact of electron-doping on the CDW phase transition, we show, in Fig. \ref{fig2}(a), $T$-dependent energy distribution curve (EDC) maps extracted from measurements such as shown in Fig. \ref{fig1}(f), as a function of $x$ (from left to right). For each $x$, lower and upper panels respectively show the RT and lowest-T EDCs. For each doping, we see the $T$-dependent energy shifts of the Se 4$p$ backfolded band, ${\Delta}E_{backf}$ (blue dots on the EDC maps), which relate to the CDW order parameter evolutions and allow for determining the transition temperatures ($T_c$) within a BCS-like scheme [onset points of the BCS-like fits in Fig. \ref{fig2}(b)] \cite{Monney2010, Mottas2019}. The impact of doping is visible on the lowest-T EDCs [Fig. \ref{fig2}(a)] with the gradual disappearance of the backfolded band with increased Ti concentration as well as on the EDCs maps which show a decrease of $T_c$, namely 217 K, 155 K, 145 K, 89 K and 62 K for our increasingly doped samples (see the red-dashed horizontal lines). 

Figure \ref{fig2}(c) shows the summary of the $T$-$x$ phase diagram of Ti-doped 1$T$-TiSe$_2$. The CDW is suppressed by Ti-intercalation with $T_c$ that linearly decreases with $x$ suggesting a quantum critical point (QCP) at $x_c$=3.55$\pm$0.18 $\%$. The observed linear dependence of $T_c$ on the Ti-concentration is well described within a Hertz-Millis picture of quantum critical scaling \cite{Hertz1976, Millis1993}, where the power-law exponent of the $x$-dependence of $T_c$ in mean-field theory is $\nu z$=1, with $\nu$ and $z$ that correspond to the exponent of the correlation length and the dynamical critical exponent, respectively. Note that our ARPES-deduced $T_c$ values closely match the ones obtained from the minima in $d\rho/dT$ in the $T$-dependent resistivity curves [see inset Fig. \ref{fig2}(c)] \cite{salvo1976}. Furthermore, our QCP value is in perfect agreement with the one estimated by Di Salvo \textit{et al.} from stoichiometry data and resistivity measurements by considering that the presence of holes is fundamental to the formation of the superlattice and that intercalated-Ti atoms increase the number of electrons and decrease the number of holes of the $p$-doped semimetallic normal state \cite{Wilson1978b, salvo1976}. 

In real-space, we can associate the decrease of $T_c$ to the recently demonstrated CDW patterning with Ti-doping \cite{Hildebrand2016}. Figures \ref{fig2}(d) shows 5.5 $\times$ 5.5 nm$^2$ STM images of the 1$T$-TiSe$_2$ surface as a function of Ti-intercalated concentration. Whereas for $x <$ 0.20 $\%$, the 2 $\times$ 2 CDW modulation can be well identified on the whole image through the presence of maxima of the charge density every second atom, in Ti-intercalated samples, DWs, where the phase of the order parameter jumps by $\pi$, proliferate (see the dotted-dashed white lines). As a function of Ti concentration, the 2 $\times$ 2 commensurate domain size shrinks and non-reconstructed 1 $\times$ 1 regions start to coexist with 2 $\times$ 2 domains, as small as the CDW coherence length, in a phase-separated pattern [see the bottom panels of Fig. \ref{fig2}(d)] \cite{Hildebrand2016}. We will now show that the spatial texturing of the CDW, characterized by the emergence of NCCDW and PS with $x$, and its suppression at the QCP both rely on the FS topology in the normal state. 

\begin{figure}[b]
\includegraphics[width=0.5\textwidth]{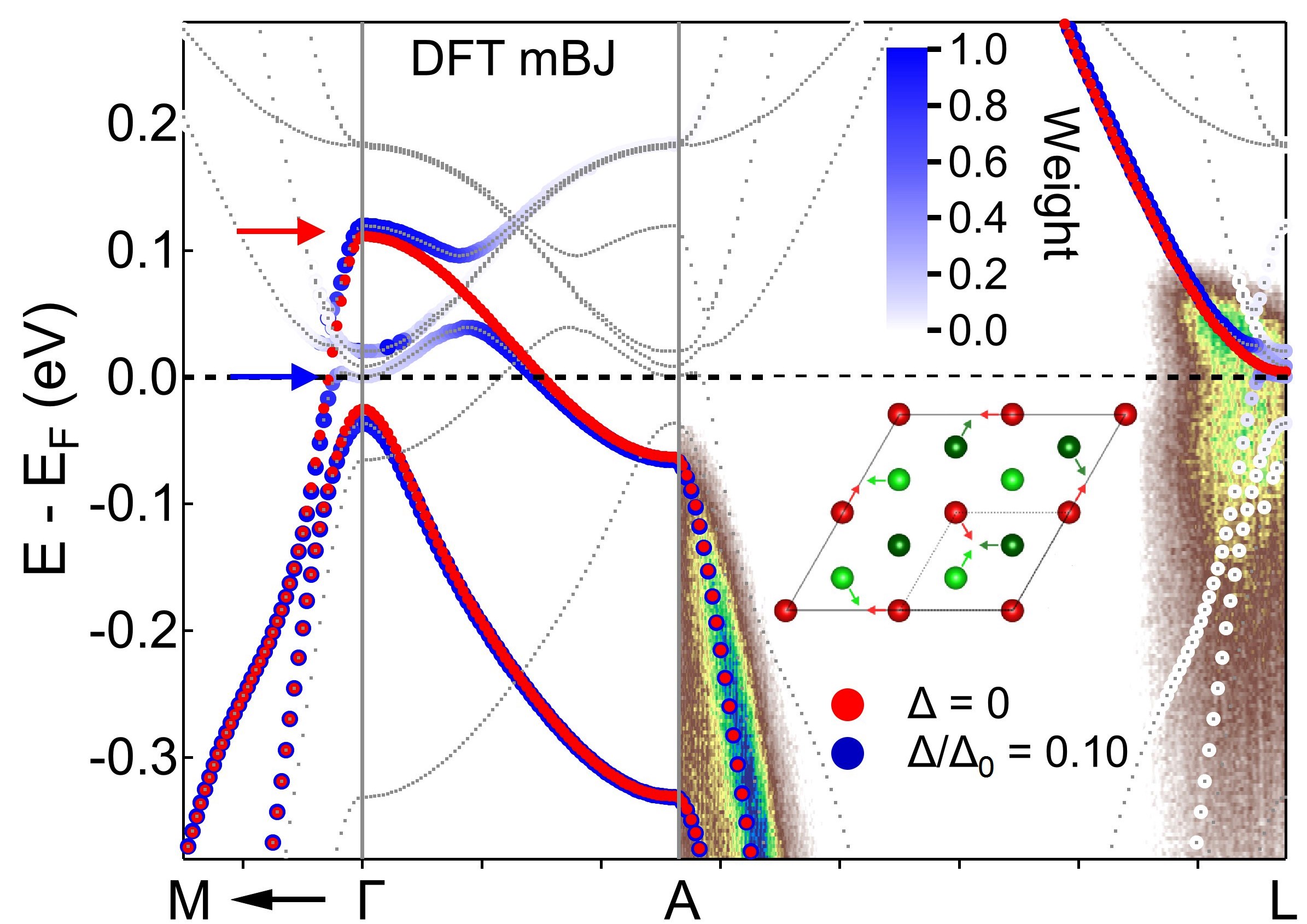}
\caption{DFT-calculated band structures of 1$T$-TiSe$_2$ along the $M$-$\Gamma$-$A$-$L$ path of the 3D BZ for the 1 $\times$ 1 $\times$ 1 normal state (red points) and of a slightly-distorted atomic structure (10 $\%$ of the full lattice distortion) according to the PLD as proposed by Di Salvo \textit{et al.} \cite{salvo1976} (blue points) and shown in the inset. The color scale from white to blue indicates the spectral weights obtained using an unfolding procedure (see Appendix B). Are also shown the 2 $\times$ 2 $\times$ 2 folded band dispersions (grey-dotted lines) and superimposed our RT ARPES measurements that mainly probe the $A$-$L$ direction of the BZ after matching the top of the calculated and experimental Se 4$p_{x,y}$ bands at $A$.}\label{fig3}
\end{figure}

\begin{figure*}[t]
\includegraphics[width=1\textwidth]{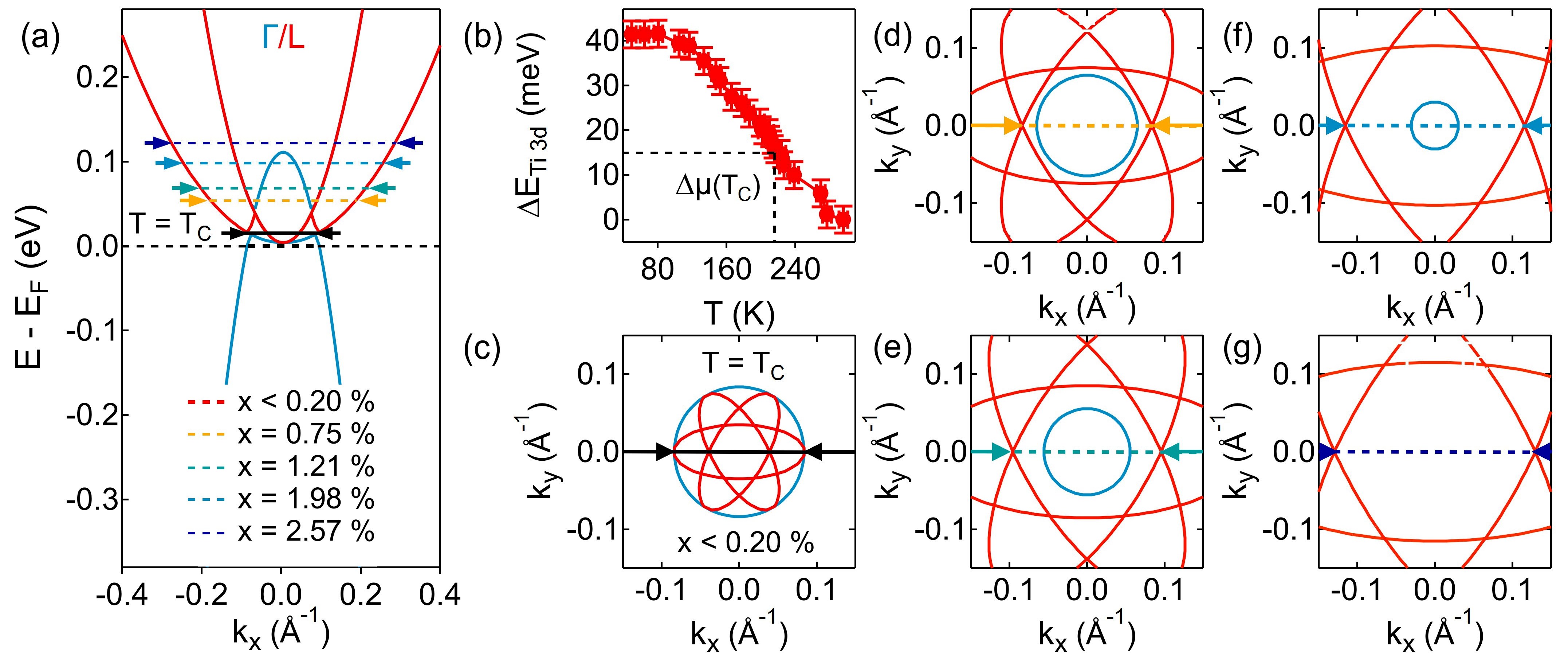}
\caption{(a) Normal-state near-$E_F$ folded dispersions of 1$T$-TiSe$_2$. The three-equivalent electron pockets (red lines, normally located at the $L$ points) are shifted here to $\Gamma$ on the hole pocket (blue line) by the three CDW $q$-vectors. The $T$-dependence of the chemical-potential shift $\Delta \mu (T)$ allows for raising $\mu$ at the electron- and hole-bands crossing at $T_c$ (black line). The coloured dashed-lines show the position of the chemical potential for the different Ti-doping extracted from ARPES. (b) $T$-dependent shift of the bottom of the Ti 3$d$ band for pristine as extracted from Fig. \ref{fig2}(a) and which relates to $\Delta \mu (T)$. (c) Sketch of the pristine 1$T$-TiSe$_2$ FS at $T$=$T_c$ and showing FS hot spots at the intersections of the folded hole (blue circle) and three-equivalent electron pockets (red ellipses). (d)-(g) Sketches of normal-state FSs for $x$=0.75 $\%$ (d), 1.21 $\%$ (e), 1.98 $\%$ (f), and 2.57 $\%$ (g). The coloured dashed-lines and arrows correspond to the $k_y$=0 \AA$^{-1}$ cuts of Fig. \ref{fig4}(a).}\label{fig4}
\end{figure*}

\paragraph*{Room-temperature pseudogap state and hidden FS}

The semimetallic or semiconducting nature of 1$T$-TiSe$_2$ still remains under debate although it has been heavily investigated in the last decades. Transport measurements \cite{salvo1976, Wilson1978b} as well as recent optical spectroscopy and conductivity studies \cite{Li2007, Velebit2016}, all concluded on a semimetallic normal state with electron-hole band overlap $\sim$-100 meV. However, three-dimensional momentum-resolved ARPES measurements have reported RT values of bandgap up to 150 meV \cite{Rossnagel2002b, Kidd2002, Zhao2007, Rasch2008, Chen2016, Watson2019}. One should recall that ARPES alone \textit{does not allow} for the quantification of the bare gap of the non-distorted phase in 1$T$-TiSe$_2$. It probes the "pseudogap" state as manifested in all ARPES studies by the presence of diffuse backfolded band intensities and suppression of spectral weight originating from CDW fluctuations already present at RT \cite{Monney2015, Borisenko2008, Pillo1999}.

The situation becomes evident using DFT calculations (see Appendix B). They reveal the hidden FS topology of the normal state and the dramatic effects of CDW fluctuations on the topmost hole-band at $\Gamma$. Figure \ref{fig3} shows DFT-calculated band structures of 1$T$-TiSe$_2$ along the $M$-$\Gamma$-$A$-$L$ path of the 3D BZ for the 1 $\times$ 1 $\times$ 1 normal state (red band structure) and an only slightly distorted structure (grey-white-blue band structure) with respect to the atomic displacements of the 2 $\times$ 2 $\times$ 2 periodic lattice distortion (PLD, shown in the inset) as determined by Di Salvo \textit{et al.} \cite{salvo1976}, and calculated for 10 $\%$ of the full PLD. This slight displacement mimics the dynamic umklapp effects associated with the Kohn anomaly, occuring above $T_c$ \cite{Holt2001a, Weber2011b}. We observe the appearance of the folded band structure because $\Gamma$, $A$ and $L$ become symmetry-equivalent (grey dotted bands) while the new spectral weight (see the color scale, from white to blue) mainly remains concentrated close to the unreconstructed (red) dispersion \cite{Voit2000}.

Also superimposed on Fig. \ref{fig3} are the RT ARPES measurements that mainly probe the $A$-$L$ direction of the BZ. The top of the calculated Se 4$p_{x,y}$ bands has been aligned with the experimental one at $A$ by shifting DFT by 24 meV. DFT very well reproduces not only our measured in-plane band dispersions but also those reported along $A$-$L$ in the most recent $k_z$-resolved ARPES studies \cite{Chen2016, Watson2019}. However we see that CDW fluctuations have a strong impact on the $\Gamma$-$A$ $k_z$-dispersion that becomes strongly renormalized as indeed measured in Refs. \cite{Chen2016, Watson2019}. This occurs due to the dynamical folding of the CDW band structure and pseudogap openings between the hybridized Se 4$p$ and Ti 3$d$ bands (also manifested at $L$ by the diffuse backfolded Se 4$p$ bands coming from $\Gamma$).

Given the $k_z$ dispersion of the topmost hole band along $\Gamma$-$A$ in the non-distorted phase, the DFT-calculated 1 $\times$ 1 $\times$ 1 state is semimetallic with an electron-hole band overlap of -107 meV defined as the energy difference between the top of the hole pocket at $\Gamma$ [red arrow in Fig. \ref{fig3}]  and the bottom of the electron pocket at $L$. Note that the out-of-plane dispersion of the hole band is crucial for the 3D component of the CDW $q$-vectors that give rise to the 2 $\times$ 2 $\times$ 2 superlattice. At RT, the CDW fluctuations hide the true 1$T$-TiSe$_2$ normal state for ARPES, which rather probes a "pseudogap" state (i.e. the blue band structure) with the bottom of the pseudogap at $\Gamma$ that corresponds to the top of the hole pocket [blue arrow on Fig. \ref{fig3} along the $\Gamma$-$M$ path of the BZ] reported in the ARPES experiments until now.  

\begin{figure*}[t]
\includegraphics[width=1\textwidth]{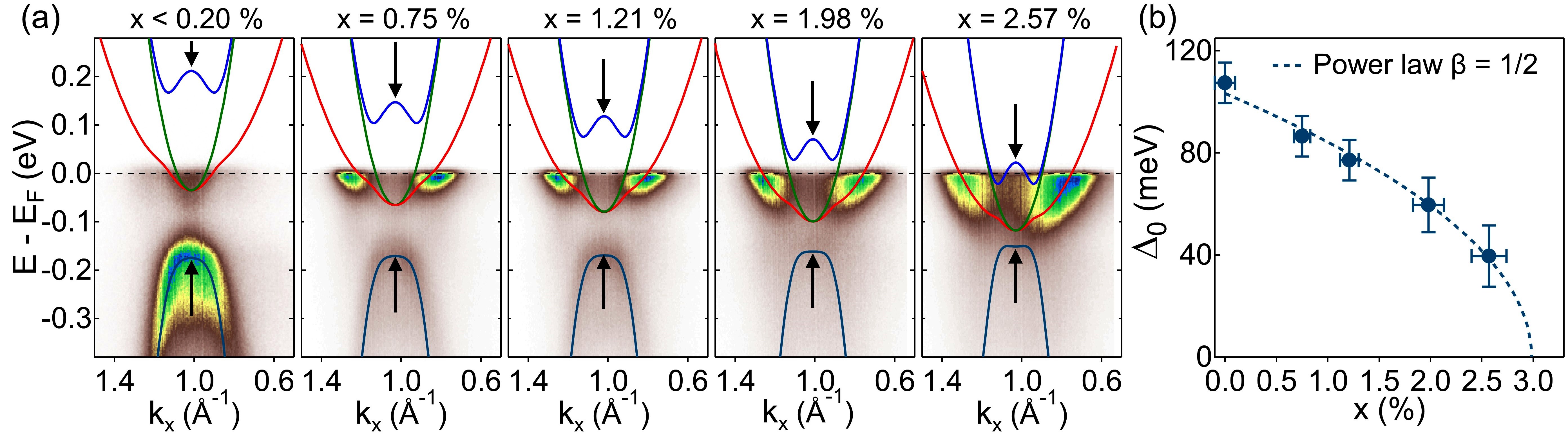}
\caption{(a) LT ARPES spectra evolution as a function of $x$ at $\bar{M}$. Are shown, in full lines, near-$E_F$ dispersions of the 1$T$-TiSe$_2$ CDW state calculated within an effective Hamiltonian model of four interacting bands with the LT order parameter $\Delta_0$ matching the experiments (see Appendix C). (b) Mean-field evolution of $\Delta_0$ as a function of the Ti concentration $x$. The blue-dashed line is a fit to the $\Delta_0$ values with $\Delta_0/\Delta_0{(x=0)}$=$(1-x/x_c)^\beta$ with $\Delta_0{(x=0)}$=103$\pm$3 meV, $\beta$=0.5 and $x_c \sim$2.99$\pm$0.15 $\%$.}\label{fig5}
\end{figure*}

\paragraph*{Fermi surface hot spots}

Let us now start with the 1$T$-TiSe$_2$ semimetallic normal state as obtained from our combined ARPES and DFT analysis [Fig. \ref{fig4}(a)]. To facilitate the discussion, we superimpose the dispersions of the hole pocket (located at $\Gamma$, blue line) and the ones of the three-equivalent electron pockets (located at the $L$ points, red lines) folded to $\Gamma$ by the three CDW $q$-vectors. Note that the lightest electron band is made of two of the symmetry-equivalent elliptical electron pockets. 

At RT, $\mu$ is placed near the lowest electron-hole band crossing (exactly 15 meV below). It raises up within the bandstructure by lowering the temperature due to the effect of the Fermi-Dirac cutoff on the overlapping hole and electron bands with different effective band masses. This temperature-dependence can be tracked in $T$-dependent ARPES measurements through the shift of the bottom of the Ti 3$d$ band at $L$ [red dots in Fig. \ref{fig2}(a), left panel] \cite{Monney2010}. Interestingly, this shift is such that, at $T$=$T_c$, $\mu$ closely matches the lowest electron- and hole-band crossing as predicted in a recent theoretical paper \cite{Chen2018} [see Fig. \ref{fig4}(b)]. 

As discussed in Ref. \cite{Monney2012}, no true nesting (i.e. connection of large parallel segments of the FS via the CDW $q$-vectors)  is necessary for driving the CDW transition in 1$T$-TiSe$_2$. Rather, as confirmed here and depicted Fig. \ref{fig4}(c), FS hot spots at the \textit{crossings} of the electron-and hole-bands (see arrows which point at two of the six crossing points) produce the instability with a wave vector corresponding to the commensurate CDW. The relevance of the hot spot physics is also corroborated by a recent optical conductivity study \cite{Velebit2016}. It has been not only demonstrated that the non-metallic $T$-dependence of transport properties in the high-temperature semimetallic phase comes from the anomalous $T$-variation of the hole and electron scattering rates, but has also reported an equal contribution of the two types of carriers to the conductivity near $T_c$ due to dominant intervalley scattering in the CDW precursor regime. 

\begin{figure*}[t]
\includegraphics[width=1\textwidth]{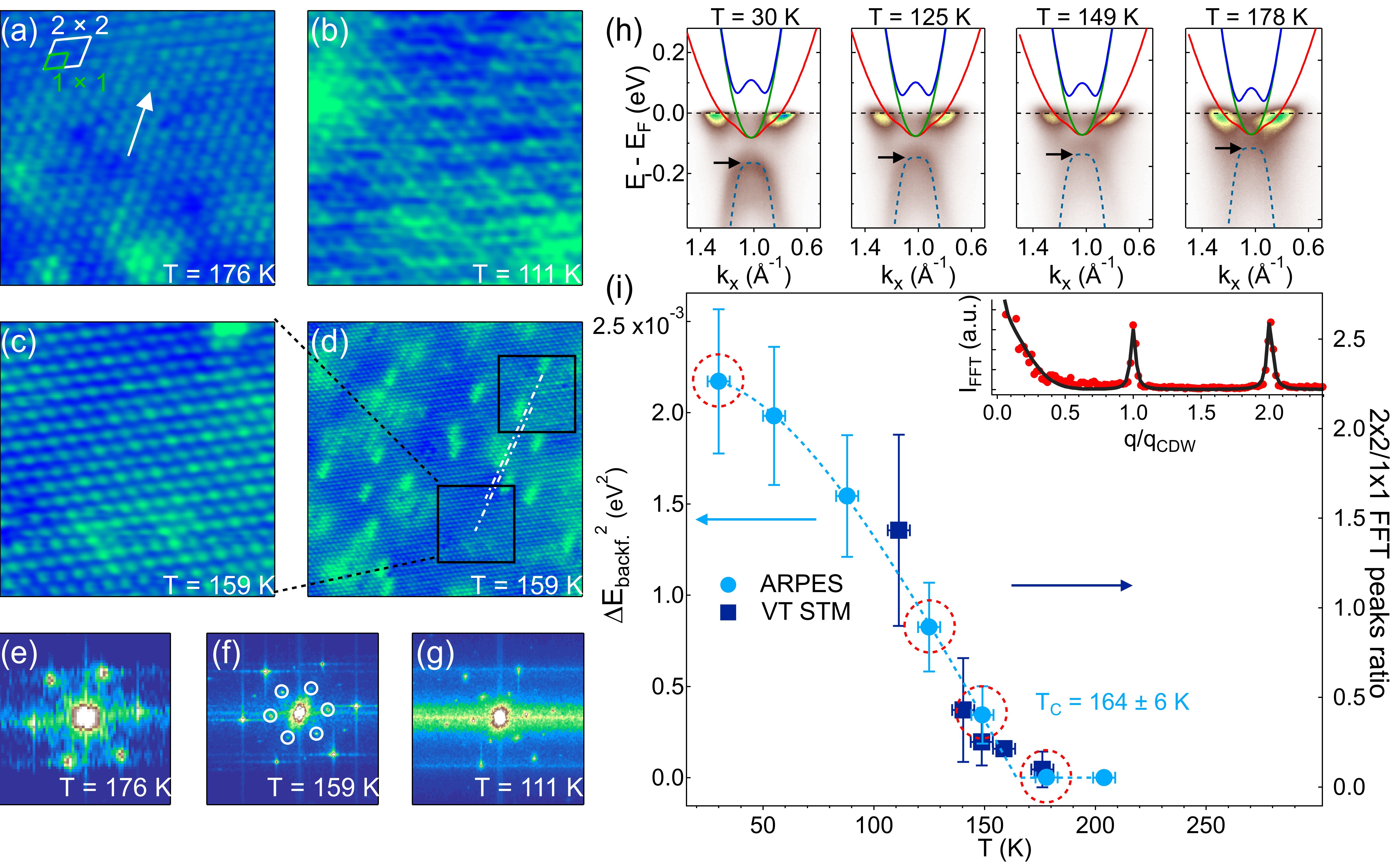}
\caption{(a)-(b), 5.5 $\times$ 5.5 nm$^2$ STM images above ($T$=176 K) and below ($T$=111 K) $T_{c}$; $V_{\text{bias}}$=50 mV, $I$=2 nA (a) and 1 nA (b). The white arrow on (a) points to a CDW droplet. The white and green rhombi indicate 2 $\times$ 2 and 1 $\times$ 1 surface unit cells. (c), 5.5 $\times$ 5.5 nm$^2$ zoom-in of a 20 $\times$ 20 nm$^2$ STM image (d) close to $T_{c}$; $V_{\text{bias}}$=50 mV, $I$=1 nA. The white-dashed lines highlight that CDW domains located by the two black squares are phase-shifted. (e)-(g), FFT-amplitude plots obtained from (a), (b) and (d). The white circles on (f) show the extra spots originating from the CDW. (h), ARPES spectra evolution at $\bar{M}$ across the CDW phase transition. (i), BCS-like $T$-dependence of ${\Delta}E_{backf}$ and $2 \times 2$/$1 \times 1$ FFT spot ratios as obtained using ARPES and VT-STM. The blue-dashed line is BCS-type gap fit to the ${\Delta}E_{backf}$ values and gives a $T_c$ of 164 $\pm$ 6 K. The red-dashed circles correspond to the ARPES spectra of (h).}\label{fig6}
\end{figure*}
  
With electron doping, we immediately see that the electron-hole band crossings move below $E_F$ [see the coloured dashed lines and arrows in Fig. \ref{fig4}(a)] and that a Lifshitz transition characterized by the loss of the hole pocket from the FS is expected close to a critical doping associated with the QCP [Fig. \ref{fig4}(d)-(g)]. Also, our LT-STM measurements revealed that with increased intercalated-Ti doping, the system undergoes a series of phase transitions from an homogeneous state to a phase-separated one, passing through a NCCDW regime [see Fig. \ref{fig2}(d)]. 

\paragraph*{Imperfect nesting}
Imperfect nesting mechanisms, based on a model of Rice \cite{Rice1970}, have been proposed for explaining the emergence of inhomogeneous electronic states upon doping in various materials such as chromium alloys, iron-based superconductors, or doped graphene bilayers \cite{Rakhmanov2013, Rakhmanov2013}. In such models, the nesting of an electron- and a hole-band is imperfect in the sense that both bands have spherical (i.e. almost parallel) FSs with slightly different radii whose mismatch is proportional to the doping \cite{Rakhmanov2017}. In this context, it has been shown that the uniform ground state is unstable with respect to PS in a wide portion of the ($x$, $T$) plane, because it becomes more favourable for the system to break up into two phases with the better or worse nesting and having different densities of itinerant electrons \cite{Rakhmanov2017}. Furthermore, depending on the level of doping and the value of $\Delta \mu$ with respect to the order parameter, it has been shown that either the system forms DWs where the doped charge is accumulated and inside which the gap vanishes locally (low doping) \cite{Gor'kov2010, Burkhardt1994}, or hosts a phase-separated state (high doping).

Therefore, since nesting is replaced by FS hot spots for driving the CDW transition in 1$T$-TiSe$_2$, we propose, in analogy with the imperfect nesting model, that the uniform CCDW ground state is unstable with respect to PS in the vicinity of these FS hot spots. While the increase of doping moves the electron-hole band crossings below $E_F$, locally, FS hot spots still exist due to PS accompanied by charge inhomogeneity. In the spatial regions with lower charge carrier densities, the CDW phase is still promoted whereas regions with higher charge carriers densities are either DWs in the NCCDW regime or non-reconstructed 1 $\times$ 1 domains in the phase-separated state on the verge of the QCP, where the Lifshitz transition in the normal state occurs \cite{Bianconi2015}.
   
\paragraph*{Doping-induced CCDW-ICDW transition}

At LT, the evolution of the 1$T$-TiSe$_2$ CDW phase upon Ti doping is shown in Figures~\ref{fig5}(a) at the $L$ point. Near-$E_F$ dispersions calculated within an effective Hamiltonian model of four bands (see Appendix C), interacting through the LT CDW order parameter $\Delta_0$, are also superimposed. While the CDW suppression with doping is evident through the suppression of the backfolded CDW band spectral weight and the closing of the CDW gap [see the black arrows in Fig. \ref{fig5}(a)], the mean-field-like evolution of $\Delta_0$ with $x$ [Fig. \ref{fig5}(b)] is in perfect agreement with the theory of the CCDW-ICDW transition for 1$T$-TiSe$_2$ upon doping as recently developed within a McMillan's Ginzburg-Landau framework \cite{Chen2019, McMillan1975, McMillan1976}. Indeed, the \textit{lock-in} energy, which controls the umklapp energy gain of locking to the commensurate structure, is proportional to the square of $\Delta_0$ \cite{salvo1976} and is closely linked to the electron density ($\propto x$ in a rigid-band model)\cite{Chen2019}. As a consequence, $\Delta_0$ is expected to vary with $\sqrt{x}$ as in Fig. \ref{fig5}(b). Moreover, it has been predicted that a Coulomb-frustrated 2D charged system hosting a second-order phase transition with an order parameter $\eta \sim \sqrt{x}$ undergoes a series of different PSs when the temperature decreases \cite{Mamin2018, Kabanov2009}, which we will address via $T$-dependent STM in the following.

\paragraph*{Phase separation around $T_c$}

As an ultimate confirmation of the mechanism at work, we thus performed VT-STM measurements of lightly-doped ($x$=0.57 $\pm$0.07 $\%$, $T_c$=164$\pm$6 K) 1$T$-TiSe$_2$ to probe the real-space emergence of the CDW with $T$. Figures \ref{fig6}(a)-(b) show 5.5 $\times$ 5.5 nm$^2$ STM images, above and below $T_c$ as obtained using ARPES. Whereas at $T$=111 K, the CDW is fully present [see Fig. \ref{fig6}(b)], interestingly, we see on Fig. \ref{fig6}a) that for temperatures higher than $T_c$ local CDWs have developed (white arrow). Approaching $T_{c}$ [Fig. \ref{fig6}(c)-(d)], we can further recognize the typical landscape of nanoscale PS with phase-shifted CDW domains embedded in the non-distorted 1 $\times$ 1 phase, as in the case of stronger doped crystals at the lowest temperatures \cite{Hildebrand2016}. The CDW domains have droplike morphologies on the scale of tens of nanometers with sharp boundaries indicating strong surface energies, which is also consistent with a Coulomb-controlled mechanism of PS \cite{Ortix2008}. The temperature evolution of the ($2 \times 2$)/($1 \times 1$) Fast-Fourier-Tranform (FFT) spot ratios [Fig. \ref{fig6}(e)-(g)] is further found to closely follow the mean-field $T$-dependence of the order parameter $\Delta$ extracted from ARPES [see arrows in Fig. \ref{fig6}(h)], reflecting the increase of the \textit{lock-in} energy upon cooling. Our VT-STM measurements thus reveal the inhomogeneous nature of the CDW emergence, the static origin of the folding of the Se 4$p$ bands visible just above $T_c$ in ARPES [right panel Fig. \ref{fig6}(h)], and demonstrate that even at such a low Ti-doping, the $T$-evolution of the CDW towards a long-range ordered state at LT intrinsically occurs through the succession of PS and NCCDW. 

It should be mentionned that quenched disorder can alter purely Coulomb-frustrated electronic configurations \cite{Campi2015, Dagotto2003}. Looking at the real-space CDW pattern morphology Fig. \ref{fig6}(d) and \ref{fig2}(d), we notice that the electron-rich bright regions locating the Ti-intercalated atoms are spatially anticorrelated with the CDW domain configuration. Unlike Cu in superconducting 1$T$-Cu$_x$TiSe$_2$ crystals \cite{Spera2019}, Ti-intercalation not only provides electrons to the Fermi sea, but also induces localized 3$d$ impurity states close to $E_F$ responsible for the formation of Ti-Ti-Ti covalent centers, strong local distortions the TiSe$_6$ octahedra \cite{Hildebrand2017}, and CDW domains pinning. Therefore, given the strong similarity between the \textit{electronic} impact of Cu and Ti on the CDW when intercalated within the 1$T$-TiSe$_2$ vdW gap \cite{Qian2007, Zhao2007}, an open question remains as to the extent to which the disorder introduced by native low-concentration Ti defects is detrimental for the emergence of superconductivity in Cu-intercalated crystals. In any case, it calls for considering an additional self-doping axis in the associated phase diagram \cite{Kogar2017} and opens the door for further exploration.  

\paragraph*{Conclusion}

We combine ARPES, LT- and VT-STM to obtain complementary momentum- and real-space insights into the CDW phase transition of the prototypical TMDC 1$T$-TiSe$_2$ natively containing low concentrations of Ti intercalants. We report persistent nanoscale PS in an extended region of the phase diagram and demonstrate that it relies on the $T$- and $x$-dependences of $\mu$ and the related FS topologies. Our work sheds light on intrinsic impurities as previously overlooked degrees of freedom for the understanding of inhomogeneous electronic landscapes associated to the emergence of superconductivity such as in 1$T$-TiSe$_2$ upon Cu intercalation or pressure. More generally, through the generic case of 1$T$-TiSe$_2$, our study extends the concept of imperfect nesting and induced electronic phase texturing to a wide class of TMDCs. 

\section{Appendix A: Experiment}
The 1$T$-TiSe$_2$ single crystals were grown at 590 \celsius , 700 \celsius , 770 \celsius , 860 \celsius , and 900 \celsius~ by iodine vapor transport, therefore containing increasing concentrations of Ti doping atoms \cite{salvo1976}. The Ti concentrations are $<$0.20 $\%$, 0.75$\pm$0.1 $\%$, 1.21$\pm$0.14 $\%$, 1.98$\pm$0.15 $\%$ and 2.57$\pm$0.22 $\%$, respectively, as determined by STM \cite{Hildebrand2016}. The low-temperature (4.5 K), respectively, variable-temperature constant current STM images were recorded using an Omicron LT-STM and VT-STM with bias voltage $V_{\text{bias}}$ applied to the sample, after cleaving \textit{in-situ} below 10$^{-7}$ mbar at room temperature. The base pressure during experiments was better than 5 $\times$ 10$^{-11}$ mbar. The temperature-dependent ARPES measurements were carried out using a Scienta DA$30$ photoelectron analyzer with He-I radiation as excitation source ($h\nu$=$21.22$ eV, SPECS UVLS with TMM 304 monochromator). The total energy resolution was 12 meV.

\section{Appendix B: Computational method}

The DFT-calculated unfolded band structure of the 2 $\times$ 2 $\times$ 2 CDW phase has been performed within the the WIEN2K package \cite{Wien}, using the modified Becke-Johnson (mBJ) exchange potential in combination with local density approximation (LDA) correlation \cite{Tran2009, Koller2012}, and including spin-orbit coupling. The parameter $c$ in mBJ was fixed for all calculations at 1.40, value determined to give the best agreement with the measured normal-state band structure of the pristine 1$T$-TiSe$_2$ and a minimum of the total energy for atomic displacements corresponding to the PLD as proposed by Di Salvo \cite{salvo1976}. The system was modeled using a 2 $\times$ 2 $\times$2 superstructure of 8 unit cells of TiSe$_2$ with lattice parameters set to $a$=$b$=3.54 \AA \space and $c$=6.01 \AA \space \cite{salvo1976}. The calculated band structures were unfolded using the FOLD2BLOCH package \cite{Rubel2014}. 

\section{Appendix C: Band simulations}

The model from which the near-$E_F$ dispersions describing ARPES of self-doped 1$T$-TiSe$_2$ have been computed includes three dimensional and anisotropic electronic bands, which are obtained by fitting the ARPES dispersions at room temperature and considering DFT band structure calculations along $k_z$. It has been shown that the electronic structure of 1$T$-TiSe$_2$ in the reconstructed phase can be well described by a Hamiltonian that has the following matrix form \cite{Monney2009a},
\begin{equation}
H=
\begin{pmatrix}
\varepsilon_v(\vec{k}) & \Delta & \Delta & \Delta \\
\Delta & \varepsilon_{c,1}(\vec{k}) &0 & 0 \\
\Delta & 0 & \varepsilon_{c,2}(\vec{k}) & 0 \\
\Delta & 0 & 0 & \varepsilon_{c,3}(\vec{k})
\end{pmatrix},\nonumber
\end{equation}
where $\Delta$ is the order parameter and describes the coupling strength between a single valence band $\varepsilon_v(\vec{k})$ at $\Gamma$ and the three symmetry equivalent conduction bands $\varepsilon_{c,i}(\vec{k})$ ($i=1,2,3$) at the three $L$ points. The band dispersions for the normal state have been chosen of the form
\begin{eqnarray}
\epsilon_v(\vec{k})&=&\hbar^2\frac{k_x^2+k_y^2}{2m_v}+t_v\cos\left(\frac{2\pi k_z}{2k_{\Gamma A}}\right)+\epsilon_v^0,\nonumber\\
\epsilon_c^i(\vec{k})&=&\frac{\hbar^2}{2m_L}\left((\vec{k}-\vec{w}_i)\cdot \vec{e}_{i\parallel}\right)^2+\frac{\hbar^2}{2m_S}\left((\vec{k}-\vec{w}_i)\cdot \vec{e}_{i\perp}\right)^2\nonumber\\&&+t_c\cos\left(\frac{2\pi (k_z-w_{iz})}{2k_{\Gamma A}}\right)+\epsilon_c^0,
\end{eqnarray}
which describe well the bands near their extrema as they are measured in ARPES experiment. The unit vectors $\vec{e}_{i\parallel}$ and $\vec{e}_{i\perp}$, pointing along the long and short axis of the ellipses, respectively, form a local in-plane basis for the electron pockets at the different $L$ points. Thus, $\vec{e}_{i\parallel}=\vec{w}_{i\parallel}/||w_{i\parallel}||$ and $\vec{e}_{i\perp}=\vec{w}_{i\perp}/||w_{i\perp}||$ where the CDW wave vectors are $\vec{w}_{i\parallel}=(w_{ix},w_{iy},0)$ and $\vec{w}_{i\perp}=(0,0,1)\times \vec{w}_i$. The $m_v$, $m_L$ and $m_S$ are the effective masses of the valence band holes and of the conduction band electrons along the long and short axis of the electron pockets, respectively. The hopping parameters $t_v$ and $t_c$ represent the amplitudes of the dispersions perpendicular to the surface and $k_{\Gamma A}$ is the distance in reciprocal space between $\Gamma$ and $A$. The parameters $\epsilon_v^0$ and $\epsilon_c^0$ are the band extrema. We define a valence band maximum $\epsilon_v^0$ lying at 110 meV above $E_F$ at the $\Gamma$ point, a $k_z$ dispersion $t_v$ of 43 meV, Ti 3$d$ conduction band minima at $L$ 4 meV above $E_F$ and a $k_z$ dispersion $t_c$ of 30 meV, yielding a semimetallic band structure with a gap $E_g$ of -107 meV (i.e. a band overlap of 107 meV). The band effective masses $m_v$, $m_L$ and $m_S$ have been fixed to their experimental values as obtained from parabolic fits of the MDCs maxima extracted from the ARPES spectra measured at RT. The effective masses values were -0.28 $\pm$0.02, 2.5$\pm$0.5, and 0.43 $m_e$ respectively where $m_e$ is the free electron mass. Note that the effective masses coming from DFT are -0.3 and 2.2 $m_e$ for the hole and electron pockets, respectively. Once the coupling $\Delta$ between the valence and conduction bands takes a finite value, the valence band gets backfolded to $L$ and the conduction bands get backfolded to $\Gamma$ and other $L$ points. The positions of the calculated four bands are complicated functions of $\Delta$ which nevertheless drastically simplify exactly at $L$ giving 
\cite{Monney2009a}, 
%\begin{eqnarray}
%&\epsilon_v(\Delta)&=\epsilon_c^0-\frac{E_g}{2}-\frac{1}{2}\sqrt{E_g^2+12\Delta^2},\nonumber \\
%&\epsilon_c^1(\Delta)&=\epsilon_c^2=\epsilon_c^0 ,\nonumber \\
%&\epsilon_c^3(\Delta)&=\epsilon_c^0-\frac{E_g}{2}+\frac{1}{2}%\sqrt{E_g^2+12\Delta^2}. 
%\end{eqnarray}
Therefore, in a photoemission experiment, we can extract the order parameter $\Delta$ from the shift of the backfolded valence band, ${\Delta}E_{backf}$ corrected by the $T$-induced chemical potential shift $\Delta \mu_T$ using,
\begin{eqnarray}
\Delta=\left(\frac{\Delta E_{backf}(\Delta E_{backf}-E_g)}{3}\right)^\frac{1}{2} 
\end{eqnarray}
with ${\Delta}E_{backf}=\epsilon_v^0 -\epsilon_v (\Delta) + E_g - \Delta \mu_T$.
\section{Acknowledgments}
This project was supported by the Fonds National Suisse pour la Recherche Scientifique through Div. II. E. Razzoli acknowledges support from the Swiss National Science Foundation (SNSF) Grant
No. P300P2-164649. We would like to thank C. Berthod and F. Weber and G. Kremer for motivating discussions. Skillful technical assistance was provided by F. Bourqui, B. Hediger and O. Raetzo.

\bibliography{library1}

\end{document}